\documentclass[sn-mathphys-num]{sn-jnl}% Math and Physical Sciences Numbered Reference Style 
\usepackage{graphicx}%
\usepackage{multirow}%
\usepackage{amsmath,amssymb,amsfonts}%
\usepackage{amsthm}%
\usepackage{mathrsfs}%
\usepackage[title]{appendix}%
\usepackage{xcolor}%
\usepackage{textcomp}%
\usepackage{manyfoot}%
\usepackage{booktabs}%
\usepackage{algorithm}%
\usepackage{algorithmicx}%
\usepackage{algpseudocode}%
\usepackage{listings}%

\algrenewcommand\algorithmicforall{\textbf{for each}}
\algrenewcommand\algorithmicrequire{\textbf{Input:}}
\algrenewcommand\algorithmicensure{\textbf{Output:}}

\begin{document}

\title[Article Title]{Hypergraphs from multivariate connectivity:\\caCOH-based EEG/MEG representation}


\author*[1]{\fnm{Daniil} \sur{Vlasenko}}\email{dvlasenko@hse.ru}

\author[1]{\fnm{Irina} \sur{Saranskaia}}\email{isaranskaya@hse.ru}

\author[1]{\fnm{Denis} \sur{Zakharov}}\email{dgzakharov@hse.ru}

\affil[1]{\orgdiv{Centre for Cognition and Decision Making}, \orgname{HSE University}, \orgaddress{\street{Myasnitskaya Ulitsa, 20}, \postcode{101000}, \state{Moscow}, \country{Russia}}}

\abstract{Hypergraphs provide a natural framework for representing neurophysiological interactions distributed across sets of sensors. A key methodological question is how hyperedges should be defined from frequency-resolved electroencephalography/magnetoencephalography (EEG/MEG) data. We demonstrate a construction strategy in which hyperedges are obtained from canonical coherence (caCOH), an extension of coherence that estimates coupling between multidimensional signal spaces. To our knowledge, this is the first work to construct hypergraphs directly from a multivariate connectivity measure specifically designed for frequency-resolved neurophysiological analysis. We propose two caCOH-based representations: a one-to-space hypergraph, where each external signal defines a hyperedge over the EEG/MEG sensor space, and a space-to-space hypergraph, where two multidimensional signal spaces are represented by a single hyperedge. We evaluate the approach in controlled simulations with known coupling frequencies and varying signal-to-noise ratio (SNR). Compared with graphs based on magnitude-squared coherence (MSC), caCOH-based hypergraphs showed statistically higher target-baseline contrasts at almost all SNR levels, indicating stronger recovery of coupling frequencies. They also recovered sensor-level spatial patterns associated with the simulated sources. In addition, one-to-space and space-to-space representations reduced 610 MSC edges per frequency to 10 hyperedges and 1 hyperedge, respectively. These results establish multivariate spectral connectivity as a natural methodological basis for EEG/MEG hypergraphs.}


\maketitle

\section{Introduction}

Multichannel neurophysiological recordings provide a way to study frequency-specific interactions between distributed neural activity and external physiological or behavioral signals. In electroencephalography (EEG)~\cite{michel_utilization_2012, cohen_where_2017} and magnetoencephalography (MEG)~\cite{hamalainen_magnetoencephalographytheory_1993, baillet_magnetoencephalography_2017}, such interactions are expressed through a set of sensors, because cortical sources project onto the broad surface of the scalp and may also be associated with external physiological or behavioral recordings, including electromyography (EMG)~\cite{luca_use_1997, conway_synchronization_1995}, electrooculography (EOG)~\cite{young_survey_1975, carter_best_2020}, or other task-related measurements. Therefore, the methodological problem is not only to detect whether coupling is present, i.e. whether signals exhibit frequency-specific statistical dependence, but also to represent its spatial and spectral structure in a compact and interpretable form. Network representations~\cite{bullmore_complex_2009, rubinov_complex_2010, bassett_network_2017} provide a natural framework for this task, because they transform connectivity estimates into structured objects that summarize interaction strength, frequency specificity, and the distribution of coupling across sensors or sources.

Classical graph approaches~\cite{berge_graphs_1973a, newman_structure_2003} provide a direct way to represent pairwise connectivity. In this framework, sensors, sources, or entire brain regions are treated as vertices, and each pairwise connectivity estimate is represented by an edge encoding the strength of interaction between the corresponding signals. In spectral analysis of EEG and MEG data, magnitude-squared coherence (MSC)~\cite{srinivasan_eeg_2007, bowyer_coherence_2016} is a standard example of such a pairwise measure: it quantifies frequency-specific linear coupling between two signals and can therefore be used to construct weighted graphs whose edges describe pairwise coupling at each frequency. However, by construction, this graph representation remains limited to pairwise interactions, because each edge is defined only for one pair of signals at a time~\cite{battiston_physics_2021, battiston_networks_2020, lambiotte_networks_2019}.

Hypergraphs~\cite{berge_graphs_1973a,  battiston_physics_2021} extend this framework by allowing a single connection, or hyperedge, to link more than two vertices. In the context of neurophysiological data, this provides a natural representation for interactions that are not confined to a single pair of sensors or sources, but are distributed across a set of signals. Instead of decomposing such interactions into many separate pairwise edges, a hypergraph can treat the whole set of involved vertices as one structured object. This makes hypergraphs particularly attractive for modeling and analyzing higher-order, i.e. set-level, organization in functional brain connectivity, where the relevant pattern may be expressed by the joint involvement of multiple sensors, sources, or regions~\cite{jie_hyperconnectivity_2016, wang_hyperedge_2018, santoro_higherorder_2024}. However, using a hypergraph as the final network model does not by itself specify what a hyperedge represents. In empirical neurophysiological data, hyperedges must be constructed from connectivity estimates, and the choice of this underlying measure determines what kind of interaction the hyperedge encodes.

The distinction between graphs and hypergraphs therefore lies in both the final network structure and the connectivity measure used to construct it. A network representation does not only summarize connectivity values; it also inherits the assumptions, strengths, and limitations of the measure from which these values are obtained. Measures based on correlation, coherence, phase synchronization, information theory, or multivariate optimization may emphasize different aspects of the same signals~\cite{david_evaluation_2004, bastos_tutorial_2016, chiarion_connectivity_2023} and can therefore lead to different network properties and interpretations~\cite{liu_benchmarking_2025, fraschini_variability_2021}. This is particularly important for hypergraph construction, because a hyperedge should ideally represent a set-level interaction rather than a post hoc grouping of already estimated pairwise links. Thus, the methodological question is not only whether to use a graph or a hypergraph, but also whether the underlying connectivity measure is consistent with the type of interaction that the chosen network model is intended to represent.

Consequently, a measure--representation mismatch arises when hypergraphs are constructed from pairwise connectivity. Pairwise measures can certainly be used to build graphs, and the resulting graphs can later be transformed into hypergraphs by aggregating edges, extracting communities, selecting neighborhoods, or applying other graph-to-hypergraph procedures (see, e.g. ~\cite{gao_hypergraph_2022, jo_edge_2021, wang_graphs_2024, li_feature_2023, wang_hyperedge_2018, yang_constructing_2023, pisarchik_hypergraph_2025, penaserrano_hypergraph_2024}). However, such transformations operate only on the information that has already been reduced to dyadic, i.e. second-order, interactions. They can reorganize pairwise links, but they do not estimate the interaction of a set of signals as a single multivariate object. The resulting hyperedges may describe collections of strong pairwise connections rather than genuinely set-level coupling patterns. In the EEG/MEG connectivity literature, a scarce number of hypergraph constructions has yet been proposed in which hyperedges are defined directly from a multivariate connectivity measure rather than obtained by aggregating pairwise links. To our knowledge, \cite{zhu_decoding_2023} is the only closely related study, but it defines multidimensional connectivity using linear regression, which is a general statistical model rather than a connectivity measure specifically designed for frequency-resolved EEG/MEG analysis. The present work addresses this gap by defining hyperedges using a connectivity measure that is both multivariate by design and naturally suited to distributed, frequency-resolved EEG/MEG interactions.

Starting from MSC as a standard pairwise coherence measure, this work focuses on canonical coherence (caCOH)~\cite{vidaurre_canonical_2019} as its multivariate counterpart. While MSC estimates frequency-specific coupling between two individual signals, caCOH extends this idea to two multidimensional signal spaces by searching for their projections whose coherence is maximal. In neurophysiological applications, these spaces may correspond, for example, to a multichannel EEG or MEG sensor space and a set of external physiological or behavioral signals. For clarity, we will subsequently consider EEG and EMG as two signal spaces. Therefore, caCOH is particularly suitable for testing the idea that hyperedges can be defined directly from multivariate spectral connectivity. In addition to the maximal coherence value, which can be used as a hyperedge weight, caCOH provides spatial patterns that indicate how individual sensors contribute to the optimized coupling. This allows the resulting hypergraph to represent both the strength of a distributed interaction and its spatial organization within the set of vertices.

In this work, we operationalize this idea by constructing and evaluating caCOH-based hypergraph representations of simulated EEG--EMG interactions. We first define MSC-based bipartite graphs as a pairwise baseline and then introduce two caCOH-based hypergraph constructions: a one-to-space variant, where each external signal defines a separate hyperedge over the EEG sensor space, and a space-to-space variant, where the interaction between two multidimensional signal spaces is represented by a single hyperedge at each frequency. For the one-to-space case, we derive an analytical solution that allows the caCOH value and the corresponding spatial pattern to be computed directly. We then test the proposed representations in controlled simulations with known frequency-specific coupling, comparing how MSC graphs and caCOH hypergraphs recover target frequencies and how this recovery depends on the signal-to-noise ratio (SNR). In addition, for caCOH-based hypergraphs, we use the scalp projections of the simulated sources as a visual reference for interpreting the recovered hyperedge spatial patterns. Together, these analyses provide a methodological demonstration that multivariate spectral connectivity can serve as a direct and effective basis for constructing hypergraph representations of distributed neurophysiological interactions.

\section{Methods}

\subsection{Pairwise connectivity: MSC-based graph representation}

As a baseline pairwise representation, we constructed  weighted complete bipartite graphs based on MSC~\cite{srinivasan_eeg_2007, bowyer_coherence_2016}. 

Let $V$ denote the set of EEG sensors, and $U$ denote the set of EMG sensors. Each EEG sensor $v_i \in V$ and each EMG sensor $u_j \in U$ are represented as vertices of a bipartite graph. Edges are allowed only between sensors of different modalities. Therefore, EEG-EEG and EMG-EMG connections are not included. For each pair of sensors $v_i$ and $u_j$, we define an undirected edge $e_{ij} \in E$ connecting them. The weight of this edge is given by the MSC between the corresponding time series $x_i(t) \in X_{EEG}$ and $y_j(t) \in Y_{EMG}$. For a given frequency $f$, MSC is defined as
\begin{equation}
\label{eq:msc}
\mathrm{MSC}_{ij}(f)
=
\frac{
\left| C_{x_i y_j}(f) \right|^2
}{
C_{x_i x_i}(f) C_{y_j y_j}(f)
},
\end{equation}
where $C_{x_i y_j}(f)$ is the cross-spectral density between the EEG signal $x_i(t)$ and the EMG signal $y_j(t)$, while $C_{x_i x_i}(f)$ and $C_{y_j y_j}(f)$ are their corresponding auto-spectral densities. Thus, the edge weight $w_{ij}(f) \in W_f$ reflects the strength of frequency-specific coupling between the two sensors and takes values between 0 and 1.

Therefore, for one EEG-EMG sample, we obtain a set of frequency-specific bipartite graphs $\{G_f\} = \{(V, U, E, W_f)\}$.

\subsection{Multivariate connectivity: caCOH-based hypergraph representation} 

While the MSC-based graph representation described above captures pairwise coupling between individual EEG and EMG sensors, it does not directly account for interactions involving groups of signals. To construct hypergraph representations, we therefore used caCOH~\cite{vidaurre_canonical_2019}, a multivariate extension of coherence designed to quantify the strongest frequency-specific coupling between two multidimensional signal spaces.

Let $A$ and $B$ denote two sets of time series, for example, EEG and EMG signals $X_{EEG}$, $Y_{EMG}$. At a given frequency $f$, the cross-spectral matrix of the joint system can be written in block form as
\begin{equation}
C(f)
=
\begin{pmatrix}
C_{AA}(f) & C_{AB}(f) \\
C_{BA}(f) & C_{BB}(f)
\end{pmatrix},
\end{equation}
where $C_{AA}(f)$ and $C_{BB}(f)$ describe the within-space spectral structure of $A$ and $B$, respectively, and $C_{AB}(f)$ describes the cross-spectral relations between the two spaces. caCOH seeks two real-valued spatial filters, $\alpha$ and $\beta$, that define linear combinations of the signals in $A$ and $B$ such that the coherence at frequency $f$ between the resulting projections is maximized. The corresponding objective function is
\begin{equation}
L(\alpha, \beta; f)
=
\frac{
    \left| \alpha^{\top} C_{AB}(f) \beta \right|^2
}{
    \left( \alpha^{\top} C_{AA}(f) \alpha \right)
    \left( \beta^{\top} C_{BB}(f) \beta \right)
}.
\label{eq:L_full}
\end{equation}
The caCOH value at frequency $f$ is then defined as the maximum of this objective over the spatial filters:
\begin{equation}
\mathrm{caCOH}^2(f)
=
\max_{\alpha, \beta}
L(\alpha, \beta; f).
\end{equation}

Thus, caCOH estimates the maximal squared absolute value of coherence that can be obtained between any linear projection of the first signal space and any linear projection of the second signal space. If both signal spaces contain only one time series, the spatial filters reduce to scalar values which cancel out, and the caCOH objective function reduces to the standard MSC formula~(eq.~\ref{eq:msc}).

In addition to the maximal coherence value, caCOH provides sensor-level weights associated with the optimized coherence at frequency $f$. After the optimal filters $\alpha$ and $\beta$ are obtained, these weights are computed as
\begin{equation}
t_{\alpha} = \operatorname{Re}(C_{AA})\alpha,
\qquad
t_{\beta} = \operatorname{Re}(C_{BB})\beta,
\end{equation}
where $\operatorname{Re}$ denotes the real part. The patterns $t_\alpha$ and $t_\beta$ are defined up to an overall sign and entries of $|t_{\alpha}|$ and $|t_{\beta}|$ indicate how strongly each sensor contributes to the maximum coherence. Therefore, in caCOH-based hypergraphs, the caCOH value can be used as a hyperedge weight, while the corresponding sensor-level weights $|t_{\alpha}|$, $|t_{\beta}|$ can be used to characterize the vertices of this hyperedge.

In the present work, we use caCOH to construct hypergraphs that represent distributed frequency-specific interactions between simulated EEG and EMG activity. In the following subsections, we describe three methodological components of this representation. First, we show that the one-to-space case of caCOH, in which coherence is estimated between a single time series and a multidimensional signal space, has an analytical solution. This contrasts with the general caCOH formulation described in the original paper~\cite{vidaurre_canonical_2019}, where the maximization problem is solved numerically. Second, we describe algorithms for using caCOH to construct hypergraphs. Third, we discuss the overfitting of caCOH and how to overcome it.

\subsubsection{Analytical solution for one-to-space caCOH} 

A particularly important special case of caCOH arises when one of the two signal spaces contains only a single time series, whereas the other space remains multivariate. This setting is common in neurophysiological applications, for example, when estimating coupling between a multidimensional EEG or MEG sensor space and a single external signal, such as one EMG sensor, a force signal, or another behavioral or peripheral measurement.

In the general caCOH formulation, the objective function involves the squared absolute value of a complex-valued coherence between two projections. Because of this form, the maximization over the spatial filters cannot be solved directly in closed form and requires computational optimization~\cite{vidaurre_canonical_2019}. In contrast, the one-to-space case has an analytical solution.

Let space $A$ contain a set of signals and space $B$ contain a single time series. In this case, the spatial filter in space $B$ reduces to a scalar factor, which cancels out in the coherence ratio (eq.~\ref{eq:L_full}). Therefore, the problem reduces to finding only the optimal spatial filter in the multivariate space $A$. It turns out, and this is demonstrated in detail in Appendix~\ref{app:one_to_space_caCOH}, that the optimal spatial filter $\alpha$ is proportional to the following expression:
\begin{equation}
\alpha
\propto
\operatorname{Re}
\left(C_{AA}(f)\right)^{-1} 
\operatorname{Re}
\left(
e^{-i\phi} C_{AB}(f)
\right),
\end{equation}
where $\operatorname{Re}$ denotes the real part, and the parameter $\phi$ is expressed analytically from the spectral matrix $C_{AA}$ and the spectral vector $C_{AB}$. In practical computations, if $\operatorname{Re}\left(C_{AA}(f)\right)$ is ill-defined, its inverse matrix can be replaced by the regularized inverse or the pseudoinverse matrix.

This analytical reduction is quite important. Instead of performing the numerical optimization used in the general caCOH setting, the one-to-space solution provides a direct way to compute quickly and accurately both the maximal coherence value and the spatial pattern $t_\alpha$ that gives rise to it.

\subsubsection{Hypergraph construction}  
The first caCOH-based construction method (Algorithm~\ref{alg:caCOH_multiple_hyperedges}) that we discuss treats each EMG sensor as a separate external signal and estimates its frequency-specific coupling with the whole EEG sensor space. For each frequency of interest $f \in F$, we construct a hypergraph $H_f$ whose vertices correspond to EEG sensors. Then, for each EMG signal $y_j \in Y_{EMG}$, we compute one-to-space caCOH between $y_j$ and the EEG signal space $X_{EEG}$. This produces two quantities: the caCOH value, which is used as the hyperedge weight, and the EEG spatial pattern, which is used as the set of vertex weights inside this hyperedge. In this representation, each EMG signal generates one hyperedge at each frequency. Therefore, a frequency-specific hypergraph contains several EMG-labelled hyperedges, each describing how strongly one EMG sensor is coupled to the distributed EEG activity at this frequency.

\begin{algorithm}
\caption{Construction of hypergraphs with one-to-space caCOH}
\label{alg:caCOH_multiple_hyperedges}
\begin{algorithmic}[1]
\Require EEG signals $X_{EEG}=\{x_1,\ldots,x_{N_{\mathrm{EEG}}}\}$, EMG signals $Y_{EMG}=\{y_1,\ldots,y_{N_{\mathrm{EMG}}}\}$, frequency set $F$
\Ensure Frequency-specific hypergraphs $\{H_f\}_{f\in F}$

\State Define EEG vertices $V_{}=\{v_1,\ldots,v_{N_{\mathrm{EEG}}}\}$
\ForAll{$f \in F$}
    \State Initialize $H_f=(V, E, W^V_f, W^E_f)$ with $E, W^V_f, W^E_f=\emptyset$
    \ForAll{EMG signal $y_j \in Y_{EMG}$}
        \State Compute one-to-space caCOH between $X_{EEG}$ and $y_j$ at frequency $f$
        \State Obtain $\mathrm{caCOH}_j^2(f)$ value and EEG spatial pattern $t_{\alpha,j}$
        \State Add hyperedge $e_{j,f} = V$ to $E$
        \State Add hyperedge weight $w^E_j(f) = \mathrm{caCOH}_j^2(f)$ to $W^E_f$
        \State Add vertex weights $w^V_j(f) = |t_{\alpha,j}|$ to $W^V_f$
    \EndFor
\EndFor

\State \Return $\{H_f\}_{f\in F}$
\end{algorithmic}
\end{algorithm}

By default, each hyperedge contains all EEG vertices. Thus, the hyperedge weight quantifies the overall strength of the EEG--EMG coupling, while the vertex weights provide a spatial pattern associated with this coupling. If a sparser representation is required, the hyperedges can be filtered using their vertex weights. For example, one may retain only the EEG sensors with the largest absolute values of $t_{\alpha,j}$. Such filtering does not change the caCOH value assigned to the hyperedge, but produces a more compact hypergraph that highlights the EEG sensors contributing most strongly to the corresponding EMG-labelled interaction. 

We used the hypergraphs constructed by Algorithm~\ref{alg:caCOH_multiple_hyperedges} to compare them with MSC-based graphs~(see Results).

The second caCOH-based construction method (Algorithm~\ref{alg:caCOH_single_hyperedge}) treats the EMG data as a multidimensional external signal space rather than as a collection of separate signals. In this case, caCOH is computed between the whole EEG signal space $X_{EEG}$ and the whole EMG signal space $Y_{EMG}$. Therefore, for each frequency $f$, we obtain one caCOH value and one EEG spatial pattern associated with the strongest coupling between these two spaces. In contrast to Algorithm~\ref{alg:caCOH_multiple_hyperedges}, this construction does not create separate hyperedges for individual EMG sensors. Instead, each frequency contributes a single hyperedge to the hypergraph.

\begin{algorithm}
\caption{Construction of hypergraph with space-to-space caCOH}
\label{alg:caCOH_single_hyperedge}
\begin{algorithmic}[1]
\Require EEG signals $X_{EEG}=\{x_1,\ldots,x_{N_{\mathrm{EEG}}}\}$, EMG signals $Y_{EMG}=\{y_1,\ldots,y_{N_{\mathrm{EMG}}}\}$, frequency set $F$
\Ensure Hypergraph $H$

\State Define EEG vertices $V=\{v_1,\ldots,v_{N_{\mathrm{EEG}}}\}$
\State Initialize $H=(V, E, W^V, W^E)$ with $E, W^V, W^E=\emptyset$
\ForAll{$f \in F$}
    \State Compute space-to-space caCOH between $X_{EEG}$ and $Y_{EMG}$ at frequency $f$
    \State Obtain $\mathrm{caCOH}^2(f)$ value and EEG spatial pattern $t_{\alpha}$
    \State Add hyperedge $e_f = V$ to $E$
    \State Add hyperedge weight $w^E(f)=\mathrm{caCOH}^2(f)$ to $W^E$
    \State Add vertex weights $w^V(f)=|t_{\alpha}|$ to $W^V$
\EndFor

\State \Return $H$
\end{algorithmic}
\end{algorithm}

By default, each hyperedge contains all EEG vertices. If needed, they can be filtered as in Algorithm~\ref{alg:caCOH_multiple_hyperedges}. Also, this algorithm can be modified to construct a bipartite hypergraph, where the second set of vertices consists of EMG sensors.

We used the hypergraphs constructed according to Algorithm~\ref{alg:caCOH_single_hyperedge} to investigate the effect of noise removal on the hyperedge weights (see the next section and Results). For the space-to-space caCOH calculation, we used the implementation provided by MNE-Connectivity~\cite{binns_mneconnectivity_2026}. This choice was appropriate because these hypergraphs were not directly compared with MSC-based graphs, and therefore did not require the same custom computational pipeline used for the MSC-based graphs and the one-to-space caCOH-based hypergraphs. At the same time, this implementation is convenient for the analysis of data rank reduction (see the next section and Results).

The two strategies can also be combined in a mixed hypergraph construction. This is useful when the external data contain several types of signals, some of which are univariate, while others form multidimensional signal spaces. In this case, each univariate external signal can be processed using the analytical one-to-space caCOH, whereas each multidimensional external signal space should be processed using the general space-to-space caCOH. 

\subsubsection{Regularization}
Because caCOH maximizes coherence over spatial filters, the resulting estimates may be sensitive to noise in high-dimensional sensor spaces. We therefore used a regularization step in both caCOH-based representations.

For our analytical one-to-space caCOH implementation, we followed the regularization strategy used in the original paper~\cite{vidaurre_canonical_2019}, which is based on singular value decomposition (SVD). At each frequency $f$, dimensionality reduction was applied to the real part of the within-space cross-spectral matrix~$\mathrm{Re}(C_{AA}(f))$. Components were ordered according to their singular values, and the smallest number of components whose cumulative singular-value sum reached 99\% of the total singular-value sum was retained. The inverse of $\mathrm{Re}(C_{AA}(f))$ in the analytical solution was then computed in this reduced subspace.

For the general space-to-space caCOH case, regularization was implemented using the rank parameter of MNE-Connectivity~\cite{binns_mneconnectivity_2026}. This parameter projects each multichannel signal space $X_{EEG}$, $Y_{EMG}$ onto a rank-reduced subspace
before caCOH is computed. The numbers of retained components were denoted by $r_{\mathrm{EEG}}$ and $r_{\mathrm{EMG}}$. For simplicity, we always used the same rank for both modalities, $r_{\mathrm{EEG}}=r_{\mathrm{EMG}}=r$. Smaller values of $r$ correspond to stronger rank reduction, whereas larger values preserve more of the original multichannel data structure. In the Results section, we vary~$r$ to examine how the retained subspace dimensionality affects the resulting caCOH hyperedge weights.

\subsection{Data simulation}
Synthetic EEG--EMG datasets were generated with MNE-Python~\cite{gramfort_meg_2013, larson_mnepython_2026a} to evaluate whether the proposed graph and hypergraph representations can recover frequency-specific coupling under controlled ground-truth conditions. In this simulation, the second modality should be interpreted more generally as a multichannel external signal driven by delayed mixtures of cortical source activity. For definiteness, we refer to this signal as EMG. The simulation was therefore designed as a controlled benchmark for recovering known EEG--external-signal coupling, rather than as a detailed biophysical model of motor-cortical or muscular activity.

Each simulated dataset consisted of $N_{\mathrm{EEG}}=61$ EEG channels and $N_{\mathrm{EMG}}=10$ EMG channels. Signals were sampled at frequency $f_s=200$ Hz for 100 s, resulting in $T=20000$ time samples per dataset. EEG sensors were arranged according to the international 10--20 system, a conventional EEG electrode-positioning scheme in which scalp electrodes are placed at standardized locations defined by relative distances between anatomical landmarks~\cite{klem_tentwenty_1999}. Cortical activity was projected to the scalp using an EEG forward model based on the fsaverage anatomy~\cite{fischl_highresolution_1999}. The forward model was computed using a realistic three-compartment volume conductor head model~\cite{nolte_analytic_2005}, with fixed source orientations. Let $N_s$ denote the number of cortical source locations and $L\in\mathbb{R}^{N_{\mathrm{EEG}}\times N_s}$ denote the resulting EEG gain matrix.

For each simulation, three ground-truth cortical source dipoles were selected randomly from the available source space. These sources were assigned independent narrowband stochastic processes in the frequency bands 10--12 Hz, 22--24 Hz, and 34--36 Hz, respectively. Each source time series was generated by band-pass filtering independent white noise with a fourth-order Butterworth filter applied in a forward--backward manner. This produced narrowband stochastic signals without introducing additional phase delays. Since the three source signals were generated independently, they were non-coherent with one another by construction.  For each simulation, we constructed a source matrix $S\in\mathbb{R}^{N_s\times T}$. Only the three selected source locations contained non-zero activity, whereas all other rows were set to zero. The clean EEG signal was then obtained as $X_{\mathrm{EEG}}^{\mathrm{sig}} = LS$.

The corresponding EMG signal was generated as a delayed multichannel mixture of the same ground-truth cortical sources. A fixed delay of $\tau=15$ ms was introduced between the cortical source activity and the external signal. This value was chosen to be within the range of previously reported EEG--EMG delays for upper-limb muscles and is close to reported forearm EEG--EMG lags of approximately 14 ms~\cite{brown_coherent_1999}. At $f_s=200$ Hz, this delay corresponds to three samples. The delayed source activity was projected to a 10-channel EMG grid arranged as $2\times5$ sensors using a randomly generated mixing matrix $M\in\mathbb{R}^{N_{\mathrm{EMG}}\times N_s}$. This matrix was generated once and reused across all simulations, so that the EMG mixing pattern remained fixed across simulation conditions. The clean EMG signal was defined as $Y_{\mathrm{EMG}}^{\mathrm{sig}}(t) = MS(t-\tau)$. Thus, the same frequency-specific cortical sources were represented in both EEG and EMG signals, but with different spatial projections and a fixed temporal offset.

Background EEG activity was generated using 500 uncorrelated cortical noise dipoles with $1/f$-type spectra. These noise dipoles were projected to the EEG sensors using the same gain matrix $L$ as the ground-truth sources. The resulting EEG noise was demeaned separately for each channel and rescaled to have a root mean square amplitude (RMS) of 10 $\mu$V. Background EMG noise was generated independently for each EMG channel as Gaussian noise. It was then demeaned separately for each channel and rescaled to have an RMS amplitude of 50 $\mu$V.

The amplitude of each simulated source signal was adjusted to obtain the target SNR. EEG simulations were performed at eight SNR levels: 0.2, 0.1, 0.05, 0.02, 0.01, 0.005, 0.002, and 0.001, whereas the EMG SNR was fixed at 0.5. SNR was computed separately for each source frequency band as the ratio between the mean channel-wise variance of the band-limited signal and the corresponding band-limited noise. Each source signal was rescaled to reach the target SNR before being added to the background noise. The final simulated EEG and EMG signals were obtained as $X_{\mathrm{EEG}} = X_{\mathrm{EEG}}^{\mathrm{sig}} + X_{\mathrm{EEG}}^{\mathrm{noise}}$ and $Y_{\mathrm{EMG}} = Y_{\mathrm{EMG}}^{\mathrm{sig}} + Y_{\mathrm{EMG}}^{\mathrm{noise}}$. For each EEG SNR level, 100 independent simulations were performed using different random seeds.

\subsection{Spectral estimation}
Spectral quantities were estimated using a windowed Fourier transform. For each simulated EEG--EMG sample, EEG and EMG signals were split into 2 s overlapping segments with 50\% overlap. At the sampling frequency $f_s=200$ Hz, this corresponded to 400 samples per segment and a step of 200 samples. Each segment was demeaned separately for each channel and multiplied by a Hann window before applying the real-valued fast Fourier transform (FFT). Frequencies from 5 to 40 Hz were retained. The resulting frequency resolution was 0.5 Hz.

For the MSC-based graphs and one-to-space caCOH hypergraphs, empirical spectral matrices were estimated by averaging products of Fourier coefficients across windows. For each frequency $f$, we computed the within-EEG spectral matrix, the within-EMG spectral matrix, and the EEG--EMG cross-spectral matrix. These matrices were then used to compute pairwise MSC edge weights and one-to-space caCOH hyperedge weights. For the space-to-space caCOH hypergraphs, caCOH was estimated with MNE-Connectivity~\cite{binns_mneconnectivity_2026} using the same 2-s overlapping epochs, FFT spectral estimation, and 5--40 Hz frequency range.

\section{Results}

\subsection{Spectral profiles of graph and hypergraph representations}

We first examined whether the MSC-based graphs and the one-to-space caCOH-based hypergraphs recover the frequency structure imposed by the simulated ground-truth sources. In each data simulation, the simulated coupling was concentrated in three narrow frequency ranges, 10--12 Hz, 22--24 Hz, and 34--36 Hz. Therefore, an appropriate representation should exhibit increased edge or hyperedge weights around these target frequencies.

\begin{figure}[!h]
\includegraphics[width=1\textwidth]{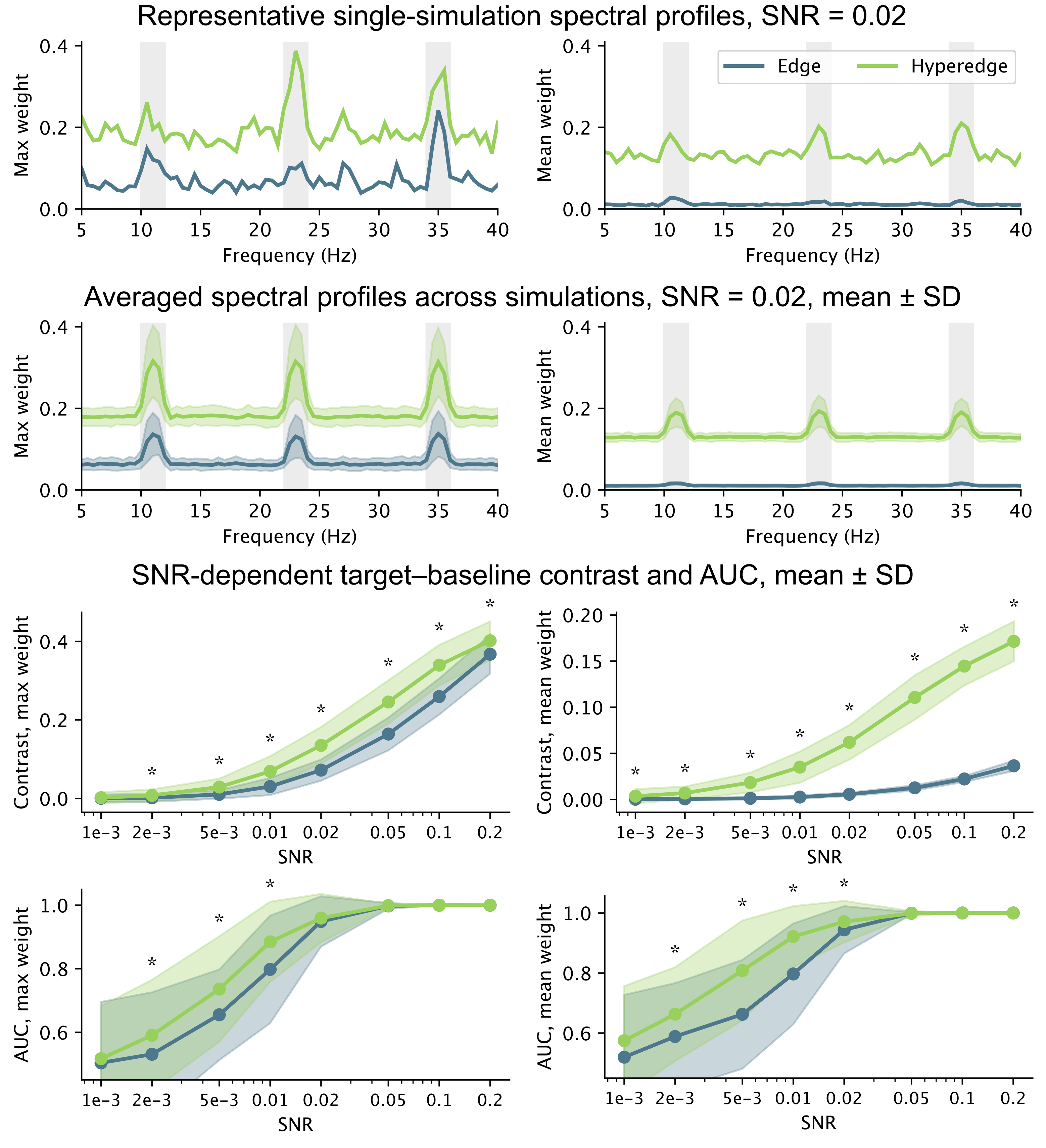}
\centering
\caption{{\bf Comparison of MSC-graphs and one-to-space caCOH-hypergraphs.} Blue curves show MSC-based graphs, and green curves show one-to-space caCOH-based hypergraphs. The left and right columns show maximum and mean connectivity weights, respectively. Shaded vertical bands indicate the simulated target-frequency ranges. \textbf{Top.} Spectral profiles for one representative simulation at an EEG SNR of 0.02. \textbf{Middle.} Spectral profiles averaged across simulations at an EEG SNR of 0.02 with standard deviation (SD). \textbf{Bottom.} SNR-dependent target--baseline contrast and AUC with SD. Asterisks (*) indicate significant differences between representations at the corresponding SNR levels.}
\label{fig:one-to-space}
\end{figure}

Fig.~\ref{fig:one-to-space} compares the spectral profiles of the MSC-based graphs and one-to-space caCOH-based hypergraphs. The first row shows profiles from one representative simulated dataset at an EEG SNR of 0.02, whereas the second row shows profiles averaged across 100 independent simulations at the same SNR level. For each frequency, the graph representation was summarized by either the maximum (left column) or the mean (right column) MSC edge weight across all EEG--EMG edges. The hypergraph representation was summarized analogously using caCOH hyperedge weights across EMG-labelled hyperedges.

The maximum-weight profiles (left column) of both representations showed local increases around the target frequency ranges, indicating that both graphs and hypergraphs were sensitive to the simulated frequency-specific coupling. However, the difference between the representations became more pronounced when weights were averaged (right column). In this case, the hypergraphs still exhibited visible target-frequency increases, whereas the corresponding graph's profile remained close to flat and showed only weak changes at the target frequencies. At the same time, the hypergraph profiles had a higher overall weight level across the full frequency range, including frequencies outside the target bands. This higher baseline is expected because, in noisy multichannel data, caCOH can enhance weak noise-related associations through spatial optimization.

\subsection{SNR-dependent recovery of target frequencies}
We next examined how the recovery of target frequencies depends on the EEG SNR. Since caCOH-based hypergraphs showed a higher overall level of weights, including at off-target frequencies, we did not compare graph and hypergraph representations only by their absolute edge or hyperedge weights. Instead, we quantified how strongly each representation separated the target frequencies from the off-target baseline.

For each simulation, representation, and spectral summary, we computed the target--baseline contrast as
\begin{equation}
\Delta =
\frac{1}{|F_{\mathrm{target}}|}
\sum_{f \in F_{\mathrm{target}}} s(f)
-
\frac{1}{|F_{\mathrm{off}}|}
\sum_{f \in F_{\mathrm{off}}} s(f),
\end{equation}
where $s(f)$ denotes either the maximum or mean edge/hyperedge weight at frequency~$f$. The set $F_{\mathrm{target}}$ consisted of the central frequencies inside the three simulated coupling ranges (11, 23, 35 Hz), whereas $F_{\mathrm{off}}$ included frequencies outside these ranges. Thus, larger values of $\Delta$ indicate a stronger separation of the simulated coupling frequencies from the background spectral level.

In addition, we computed an AUC-type frequency-discrimination measure. For each simulation, representation, and spectral summary, AUC was defined as
\begin{equation}
\mathrm{AUC}
=
\frac{1}{|F_{\mathrm{target}}||F_{\mathrm{off}}|}
\sum_{f_t \in F_{\mathrm{target}}}
\sum_{f_o \in F_{\mathrm{off}}}
\left(
\mathbb{I}\{s(f_t) > s(f_o)\}
+
\frac{1}{2}\mathbb{I}\{s(f_t) = s(f_o)\}
\right),
\end{equation}
where $\mathbb{I}$ denotes the indicator function, which equals 1 if the condition inside the braces is true and 0 otherwise. This measure can be interpreted as the probability that a randomly selected center target frequency $f \in F_{\mathrm{target}}$ receives a higher spectral score than a randomly selected off-target frequency $f \in F_{\mathrm{off}}$, with ties counted as one half. Thus, $\mathrm{AUC}=0.5$ corresponds to chance-level discrimination, whereas $\mathrm{AUC}=1$ indicates perfect separation.

The bottom two rows of Fig.~\ref{fig:one-to-space} show the SNR-dependent target--baseline contrast and AUC for the maximum-weight (left column) and mean-weight (right column) profiles. For both graph and hypergraph representations, the contrast and AUC increased with SNR, indicating that stronger simulated coupling led to clearer recovery of the target frequencies. Graph and hypergraph contrast values were compared at each SNR level using the paired Wilcoxon signed-rank test across matched simulations. Holm correction across SNR levels was applied separately for the tests based on maximum weight, mean weight, contrast, and AUC, with the significance level set to $\alpha = 0.01$ for contrast and $\alpha = 0.1$ for AUC. Significant differences are marked by asterisks (*). The caCOH-based hypergraphs showed significantly higher contrast than the MSC-based graphs in all tested conditions except the lowest SNR level (0.001) for the maximum-weight profile. Hypergraphs also showed significantly higher AUC for the maximum-weight profile at SNRs of 0.002, 0.005, and 0.01 and for the mean-weight profile at SNRs of 0.002, 0.005, 0.01, and 0.02.

\subsection{Spatial patterns recovered by caCOH-based hypergraphs}
\begin{figure}[!h]
\includegraphics[width=0.925\textwidth]{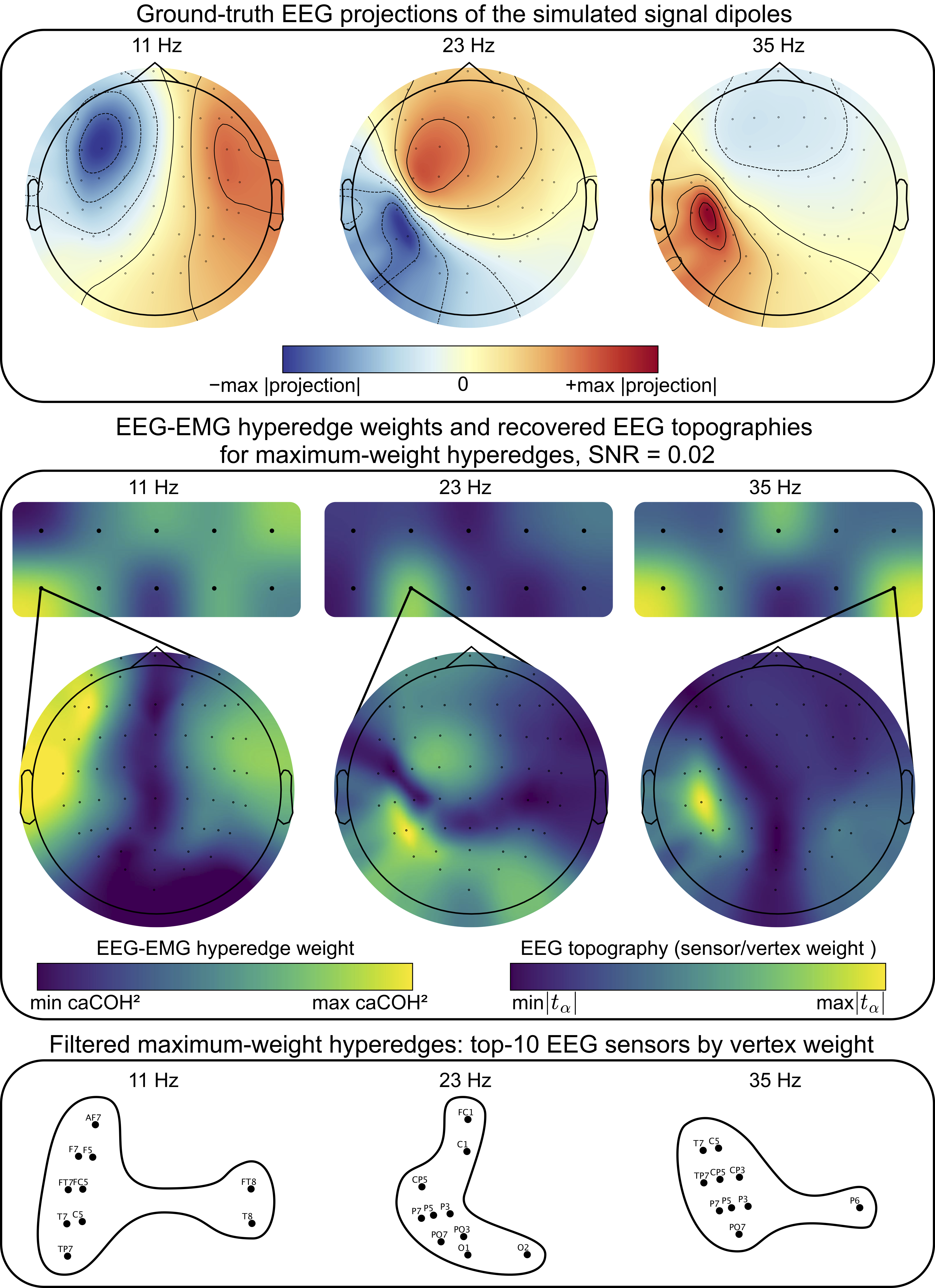}
\centering
\caption{{\bf Spatial patterns recovered by one-to-space caCOH-based hypergraphs.} The columns show the three target frequencies. \textbf{Top.} Ground-truth EEG sensor-level projections of the simulated cortical source dipoles.  \textbf{Middle.} Hypergraph representation recovered from EEG--EMG data at an EEG SNR of 0.02. EMG grids show the weights of hyperedges, whereas EEG topographies show the vertex weights of the maximum-weight hyperedges. \textbf{Bottom.} Sparsified maximum-weight hyperedges retaining the top-10 EEG sensors, providing a compact spatial representation.
}
\label{fig:hyperedges}
\end{figure}

The caCOH-based hypergraph representation provides a spatial description of the recovered EEG--EMG interactions. Each hyperedge is associated with an EMG signal, has a scalar hyperedge weight reflecting the strength of coupling, and contains EEG vertex weights that describe the spatial pattern of EEG sensors contributing to this interaction. Fig.~\ref{fig:hyperedges} illustrates this structure for the same representative simulation with an EEG SNR of 0.02, whose spectral profiles are shown in the top row of Fig.~\ref{fig:one-to-space}. 

Fig.~\ref{fig:hyperedges} demonstrates three target frequencies corresponding to the centers of the simulated coupling ranges, 11 Hz, 23 Hz, and 35 Hz. The top row shows the ground-truth EEG projections of the three simulated cortical sources. These maps represent the spatial distribution that each source induces at the EEG sensor level through the forward model.

The middle row of Fig.~\ref{fig:hyperedges} shows the hypergraph representation recovered from EEG--EMG data. For each target frequency, the EMG grid represents the weights of EMG-labelled hyperedges, whereas the EEG topography shows the vertex weights of the maximum-weight hyperedge. Even at an EEG SNR of 0.02, the recovered EEG patterns remained qualitatively close to the corresponding ground-truth projections, although minor deviations were visible. Thus, the one-to-space caCOH hypergraph recovered the main spatial structure of the simulated coupling even under substantial noise.

The bottom row of Fig.~\ref{fig:hyperedges} shows filtered versions of the same maximum-weight hyperedges, where only the 10 EEG sensors with the largest vertex weights were retained. This filtering converts a dense hyperedge into a more compact spatial representation. Such sparsification may be useful in hypergraph analyses, where mathematical transformations of hypergraphs are applied to extract structural information from the interactions.

\subsection{Effect of regularization on caCOH-based hypergraphs}
\label{res:reg}
Next, we examined how regularization affects the resulting hypergraphs using the space-to-space caCOH representation. In this case, caCOH was computed between the full EEG and EMG signal spaces, producing one hyperedge for each frequency. The rank parameter $r$ controlled the dimensionality of the retained rank-reduced subspace before caCOH estimation, with smaller values corresponding to stronger rank reduction.

\begin{figure}[!h]
\includegraphics[width=1\textwidth]{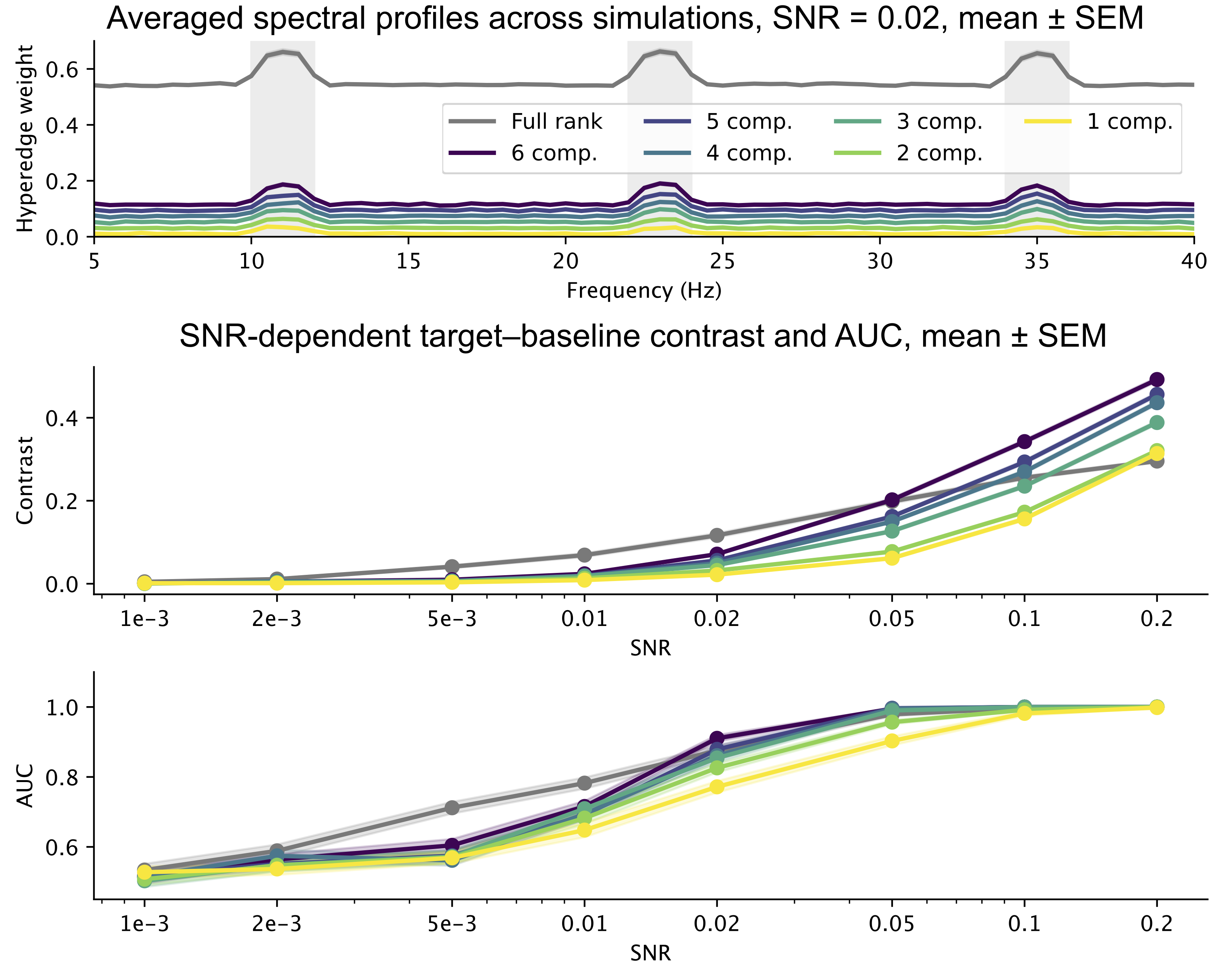}
\centering
\caption{{\bf Effect of rank reduction on space-to-space caCOH-based hypergraphs.} Curves correspond to different numbers of retained SVD components. Shaded vertical bands indicate the simulated target-frequency ranges. \textbf{Top.} Spectral profiles averaged across simulations at an EEG SNR of 0.02 with SEM. \textbf{Bottom.} SNR-dependent target--baseline contrast and AUC for different rank-reduction levels with SEM.}
\label{fig:rank-reduction}
\end{figure}

Fig.~\ref{fig:rank-reduction} shows the resulting spectral profiles (top row) and target--baseline contrasts and AUC (two bottom rows) for different rank-reduction levels. The averaged spectral profiles at the EEG SNR of 0.02 (top row) show that all variants retained peaks near the target frequencies, but the baseline level and peak sharpness depended on the retained rank. It can be observed that the less aggressive the filtering, the higher the overall spectral profile and the sharper the peaks. To make the comparison between multiple curves visually interpretable, variability is shown as standard error of the mean (SEM) rather than SD.

The SNR-dependent contrast and AUC analysis (two bottom rows) further shows that the effect of rank reduction depends on the signal strength. At the lowest SNR levels, aggressive rank reduction can remove a part of the weak but informative signal structure together with the noise. In this regime, preserving more of the original multichannel data may therefore lead to higher target--baseline contrast. As the SNR increases, rank reduction becomes more beneficial: removing low-variance or noisy directions improves the separation of target frequencies from the baseline.

\subsection{Compactness of caCOH-based hypergraphs}
The proposed hypergraph representations also provide a more compact description of connectivity than the corresponding graph representation. For each frequency, the MSC-based bipartite graph contains one edge for every EEG--EMG sensor pair. In our simulations, this gives $N_{\mathrm{EEG}} \times N_{\mathrm{EMG}} = 61 \times 10 = 610$ weighted edges per frequency. In contrast, the one-to-space caCOH-based hypergraph contains one hyperedge for each EMG signal, resulting in only $N_{\mathrm{EMG}} = 10$ weighted hyperedges per frequency. The space-to-space caCOH-based hypergraph is even more compact, because the interaction between the full EEG and EMG signal spaces is represented by a single weighted hyperedge at each frequency.

Therefore, at the level of stored connectivity objects, the one-to-space hypergraph reduces the number of links by a factor of 61 compared with the graph, whereas the space-to-space hypergraph reduces it by a factor of 610. Across a set of frequencies $F$, this corresponds to $610|F|$ graph edges, $10|F|$ one-to-space hyperedges, and only $|F|$ space-to-space hyperedges. This difference becomes especially relevant when frequency-resolved representations are stored for many subjects, time windows, or experimental conditions.

Importantly, this comparison concerns the basic connectivity structure, that is, the number of weighted edges or hyperedges required to represent the interactions. Vertex weights in caCOH-based hypergraphs are optional annotations: they can be omitted when only interaction strengths are needed or removed after hyperedge filtering when spatial interpretability is required. This compactness may benefit downstream analyses that iterate over links, compare representations across conditions, or store large collections of frequency-specific connectivity patterns. 

\section{Discussion}
This work addresses a methodological problem in EEG/MEG hypergraph analysis: how to define hyperedges corresponding to set-level interactions estimated directly from the data rather than to post hoc groupings of pairwise connectivity links. Using this principle, we introduced two hypergraph construction methods. In controlled simulations with known frequency-specific EEG--EMG coupling, caCOH-based hypergraphs separated target coupling frequencies from the off-target baseline more strongly than MSC-based graphs, while also reducing 610 pairwise EEG--EMG edges per frequency to 10 one-to-space hyperedges or to a single space-to-space hyperedge. However, our results should be viewed as a controlled methodological benchmark rather than as empirical validation. In future research, this method should be tested using experimental data.

The proposed framework may be useful whenever neurophysiological coupling is naturally distributed across sets of signals, including EEG/MEG interactions with EMG, EOG, force, movement, or other behavioral measurements, as well as analyses involving multiple time windows, experimental conditions, or hyperscanning recordings from more than one participant. Overall, this work supports multivariate spectral connectivity as a natural and promising basis for constructing hypergraph representations of frequency-resolved neurophysiological data.

\subsection{Differences between one-to-space and space-to-space approaches}
The two proposed constructions differ in the amount of external-signal information used during caCOH optimization. In the one-to-space case, each hyperedge is estimated from a single external time series, such as one EMG channel, whereas the space-to-space construction uses the full multichannel EMG space and therefore has more degrees of freedom to identify the maximally coherent EEG--EMG coupling pattern. This difference is important for interpretation. If a particular EMG channel only weakly represents the external activity coupled to the EEG space, as may happen in our simulations depending on the EMG mixing matrix $M$, its one-to-space hyperedge may have a lower weight than the space-to-space hyperedge, which can optimize over all EMG channels. In such cases, the associated EEG spatial pattern should also be interpreted more cautiously. Therefore, the hyperedge weight can be interpreted not only as coupling strength, but also as a practical indicator of the reliability of the associated EEG pattern.

This may raise the question of why one should use the one-to-space approach at all if the space-to-space approach can estimate the frequencies of maximal coupling more effectively due to its greater optimization freedom. First, in some empirical settings only one external sensor may be available. Second, the one-to-space approach is less exposed to noise in multichannel EMG data, because it cannot optimize the EEG projection with respect to noise distributed across several EMG sensors. Third, the analytical solution for the one-to-space case makes the computation faster. Fourth, the one-to-space construction provides an EMG-sensor-specific view of EEG--EMG coupling, which may be important in applications where the coupling of individual external sensors is meaningful, for example in neuroprosthetics.

\subsection{Regularization and preprocessing considerations}
As shown in Section~\ref{res:reg}, regularization can substantially affect caCOH-based hypergraphs, which raises the practical question of how the data should be preprocessed before hypergraph construction. A universal rule is difficult to provide, because the optimal strategy depends on the noise level, sensor configuration, and experimental design. In general, EEG/MEG and external recordings should first undergo standard preprocessing, including filtering, removal of power-line noise, bad-channel correction, rejection of noisy segments, and artifact correction using automated and manual procedures such as independent component analysis~(ICA)~\cite{bigdely-shamlo_prep_2015, jiang_removal_2019, gross_good_2013}. After this signal-level preprocessing, additional regularization can be applied during caCOH computation. In the one-to-space case, we followed the strategy proposed in the original caCOH study~\cite{vidaurre_canonical_2019} and applied SVD-based regularization directly to the spectral matrix, retaining 99\% of the spectral information. This second regularization step is intended to suppress residual noisy or unstable dimensions that remain after preprocessing of the original time series.

\subsection{Choice of a multivariate connectivity measure}
We used caCOH as a concrete example of a multivariate spectral connectivity measure for hyperedge construction. However, the choice of a particular measure should depend on the research question and data structure. caCOH is naturally suited to estimate coupling between two signal spaces, such as EEG/MEG activity and external physiological or behavioral measurements, but it may be less appropriate for within-space EEG/MEG connectivity, where broadly projected neural sources and volume conduction can make optimized projections reflect shared field spread rather than interactions between neural sources.

\backmatter

\section*{Statements and declarations}

\bmhead{Funding}
This research was supported by the RSF (project No. 24-68-00030) and in part through computational resources of HPC facilities at HSE University~\cite{kostenetskiy_hpc_2021}.

\bmhead{Competing interests}
The authors declare that they have no competing interests.

\bmhead{Ethics approval and consent to participate}
Not applicable.

\bmhead{Consent for publication}
Not applicable.

\bmhead{Data availability}
No empirical datasets were used in this study. The generated EEG--EMG dataset is publicly available~\cite{vlasenko_eeg_2026}.

\bmhead{Materials availability}
Not applicable.

\bmhead{Code availability}
The code used to generate the simulations and perform the analyses is available from the corresponding author upon reasonable request.

\bmhead{Author contribution}
D.V. contributed to conceptualization, methodology, software development for the MSC-based graph, caCOH-based hypergraph, and spectral-analysis pipelines, formal analysis, visualization, writing of the original draft, and project administration.
I.S. contributed to software development for the data simulation, visualization, drafting of the simulation-related description, and writing -- review and editing.
D.Z. contributed to supervision, project administration, funding acquisition, and writing -- review and editing.
All authors read and approved the final manuscript.

\begin{appendices}
\section{Analytical solution for one-to-space caCOH}
\label{app:one_to_space_caCOH}
We consider a fixed frequency $f$ and omit the explicit frequency argument for readability. The caCOH objective function is originally defined as the squared absolute value of coherence between two projected signals:
\begin{equation}
\label{eq:ap:absObj}
L(\alpha,\beta)
=
\frac{
\left|\alpha^{T} C_{AB} \beta\right|^{2}
}{
\alpha^{T} C_{AA} \alpha \, \beta^{T} C_{BB} \beta
}.
\end{equation}
Here, $C_{AB}$ is the complex cross-spectral matrix between signal spaces $A$ and $B$, whereas $C_{AA}$ and $C_{BB}$ are the corresponding within-space cross-spectral matrices. The spatial filters $\alpha$ and $\beta$ are real-valued, since they represent linear mixtures of the original signals.

The numerator contains the absolute value of a complex scalar. For any fixed pair of filters $\alpha$ and $\beta$, the complex number $\alpha^{T}C_{AB}\beta$ can be phase-rotated so that its value becomes real. This motivates the introduction of a phase parameter $\phi$. Instead of directly optimizing the absolute value, the caCOH formulation can be rewritten in terms of the real part of the phase-rotated cross-spectrum $\operatorname{Re} \left(e^{-i\phi} C_{AB}\right)$. The objective function from eq.~\ref{eq:ap:absObj} can therefore be written as
\begin{equation}
\label{eq:ap:notAbsObj}
L(\alpha,\beta,\phi)
=
\frac{
\left(\alpha^{T} \operatorname{Re} \left(e^{-i\phi} C_{AB}\right) \beta\right)^2
}{
\alpha^{T} C_{AA} \alpha \, \beta^{T} C_{BB} \beta
}.
\end{equation}
In the general caCOH formulation, the optimal phase is not known in advance, and the remaining optimization over $\phi$ is performed numerically~\cite{vidaurre_canonical_2019}. Below, we show that in the one-to-space case this phase optimization has an analytical solution.

Let space $A$ be multivariate, while space $B$ contain only one time series. Then $\beta$ is a scalar, and its contribution cancels out between the numerator and the denominator. Therefore, without loss of generality, we can set $\beta = 1$. Since space $B$ contains only one time series, $C_{AB}$ becomes a vector, and $C_{BB}$ reduces to the auto-spectrum of one signal, which is a real non-negative scalar.

We first express the optimal spatial filter $\alpha$ for a fixed phase $\phi$. For clarity, let us introduce the notation
$r(\phi)
=
\operatorname{Re}
\left(
e^{-i\phi} C_{AB}
\right)$.
Since $C_{AA}$ is a within-space cross-spectral matrix, it is Hermitian. Because the spatial filter $\alpha$ is real-valued, the imaginary part of $C_{AA}$, which is skew-symmetric, does not contribute to the quadratic form $\alpha^T C_{AA}\alpha$. Therefore, the denominator depends only on the real part of $C_{AA}$, and we define $C = \operatorname{Re}(C_{AA})$. The objective function from eq.~\ref{eq:ap:notAbsObj} then becomes
\begin{equation}
\label{eq:ap:simplObj}
L(\alpha,\phi)
=
\frac{
\left(\alpha^{T}r(\phi)\right)^2
}{
\alpha^{T} C \alpha \, C_{BB}
}.
\end{equation}

The numerator can be rewritten as
\begin{equation}
\alpha^{T}r(\phi)
=
\alpha^{T}C^{1/2}C^{-1/2}r(\phi).
\end{equation}
Here and below, we assume that $C$ is positive definite, so that $C^{-1/2}$ is well defined. In practical computations, this inverse can be replaced by a regularized inverse or pseudoinverse matrix. Introducing $u = C^{1/2}\alpha$ and $v = C^{-1/2}r(\phi)$, we obtain
\begin{equation}
\alpha^{T}r(\phi) = u^{T}v.
\end{equation}
By the Cauchy--Schwarz inequality, $(u^{T}v)^2 \leq (u^{T}u)(v^{T}v)$. Substituting $u$ and $v$ back gives
\begin{equation}
\left(\alpha^{T}r(\phi)\right)^2
\leq
\left(\alpha^{T}C\alpha\right)
\left(r(\phi)^{T}C^{-1}r(\phi)\right).
\end{equation}
Therefore,
\begin{equation}
L(\alpha,\phi)
\leq
\frac{
r(\phi)^{T}C^{-1}r(\phi)
}{
C_{BB}
}.
\end{equation}
The equality is reached when $u$ and $v$ are collinear, that is,
$
C^{1/2}\alpha \propto C^{-1/2}r(\phi).
$
Equivalently,
\begin{equation}
\alpha \propto C^{-1}r(\phi).
\end{equation}
Thus, for a fixed phase $\phi$, the optimal spatial filter in space $A$ is proportional to
\begin{equation}
\alpha_{0} = C^{-1}r(\phi).
\end{equation}
If we normalize this filter with respect to the $C$-induced norm, we can write
\begin{equation}
\alpha
=
\frac{\alpha_{0}}{\sqrt{\alpha_{0}^TC\alpha_{0}}}
=
\frac{\alpha_{0}}{\sqrt{\alpha_{0}^TCC^{-1}r(\phi)}}
=
\frac{\alpha_{0}}{\sqrt{r(\phi)^{T}\alpha_{0}}},
\end{equation}
which gives
\begin{equation}
\label{eq:simp1}
\alpha^{T}C\alpha
=
\frac{\alpha_0^TC\alpha_0}{r(\phi)^T\alpha_0}
=
1.
\end{equation}
Also, using the normalized filter, we have
\begin{equation}
\label{eq:simp2}
\left(r(\phi)^T \alpha\right)^2
=
\left(
r(\phi)^T
\frac{\alpha_0}{\sqrt{r(\phi)^T \alpha_0}}
\right)^2
=
r(\phi)^T \alpha_0
=
r(\phi)^T C^{-1}r(\phi).
\end{equation}
Using eq.~\ref{eq:simp1} and eq.~\ref{eq:simp2} in eq.~\ref{eq:ap:simplObj}, the objective function after maximization over $\alpha$ is
\begin{equation}
\label{eq:ap:maxAObj}
\max_{\alpha} L(\alpha,\phi)
=
\frac{
r(\phi)^T C^{-1}r(\phi)
}{
C_{BB}
}.
\end{equation}

It remains to maximize the objective function over the phase $\phi$. Let $C_{AB}=x+iy$, where $x$ and $y$ are real vectors. Then
\begin{equation}
e^{-i\phi}C_{AB}
=
(\cos\phi-i\sin\phi)(x+iy),
\end{equation}
and therefore
\begin{equation}
\label{eq:ap:simR}
r(\phi)
=
\operatorname{Re}
\left(
e^{-i\phi}C_{AB}
\right)
=
x\cos\phi + y\sin\phi.
\end{equation}
Substituting eq.~\ref{eq:ap:simR} into eq.~\ref{eq:ap:maxAObj} gives
\begin{equation}
\label{eq:ap:maxAObjsincos}
\max_{\alpha}L(\alpha, \phi)
=
\frac{
(x\cos\phi+y\sin\phi)^{T}
C^{-1}
(x\cos\phi+y\sin\phi)
}{
C_{BB}
}.
\end{equation}
Let's define the scalar quantities
$a=x^{T}C^{-1}x$, $b=x^{T}C^{-1}y$, $c=y^{T}C^{-1}y$, then eq.~\ref{eq:ap:maxAObjsincos} becomes
\begin{equation}
\label{eq:ap:maxAObjsincos2}
\max_{\alpha}L(\alpha, \phi)
=
\frac{
a\cos^{2}\phi
+
2b\sin\phi\cos\phi
+
c\sin^{2}\phi
}{
C_{BB}
}.
\end{equation}
Using the identities
$\cos^{2}\phi=(1+\cos 2\phi)/2$,
$\sin^{2}\phi=(1-\cos 2\phi)/2$, and
$2\sin\phi\cos\phi=\sin 2\phi$,
we turn eq.~\ref{eq:ap:maxAObjsincos2} into
\begin{equation}
\label{eq:ap:maxAObjsincos3}
\max_{\alpha}L(\alpha, \phi)
=
\frac{
\frac{a+c}{2}
+
\frac{a-c}{2}\cos 2\phi
+
b\sin 2\phi
}{
C_{BB}
}.
\end{equation}
Thus, the remaining phase-dependent part has the form $p\cos\theta+q\sin\theta$, where $p=(a-c)/2$, $q=b$, and $\theta=2\phi$. This expression can be written as $p\cos\theta+q\sin\theta = R\cos(\theta-\delta)$, where $R=\sqrt{p^{2}+q^{2}}$ and $\delta=\operatorname{atan2}(q,p)$ is the quadrant-aware angle satisfying $\cos\delta=p/R$ and $\sin\delta=q/R$. Therefore, the maximum is reached when $\theta=\delta+2\pi n$, $n\in\mathbb{Z}$. Since $\theta=2\phi$, the optimal phase is
\begin{equation}
\phi^*
=
\frac{\delta}{2} + \pi n,
\qquad
n\in\mathbb{Z}.
\end{equation}
Using this in eq.~\ref{eq:ap:maxAObjsincos3}, the maximal value of the one-to-space caCOH objective function is therefore
\begin{equation}
\begin{aligned}
\max_{\alpha,\phi} L(\alpha,\phi)
&=
\frac{1}{C_{BB}}
\left(
\frac{a+c}{2}
+
\sqrt{
\left(\frac{a-c}{2}\right)^2+b^2
}
\right) \\
&=
\frac{1}{2C_{BB}}
\left(
a+c+\sqrt{(a-c)^2+4b^2}
\right).
\end{aligned}
\end{equation}

Thus, when space $A$ is multivariate and space $B$ contains only a single time series, caCOH has an analytical solution.

\end{appendices}

% \bibliography{MSCGvscaCOHHG.bib}
\bibliography{Hypergraphs.bib}

\end{document}